\documentclass[twocolumn,prd,groupedaddress,nofootinbib,showpacs]
{revtex4}

\usepackage{latexsym}% D'Alambertian operator \Box
\usepackage{amssymb}% D'Alambertian operator \square
\usepackage{graphicx}% Include figure files
\usepackage{dcolumn}% Align table columns on decimal point
\usepackage{bm}% bold math

\begin{document}

\title{Noncommutative fields in curved space}

\author{J. Barcelos-Neto}
\email{barcelos@if.ufrj.br}
\affiliation{Instituto de F\'{\i}sica\\
Universidade Federal do Rio de Janeiro\\
Caixa Postal 68528, RJ 21941-972 -- Brazil}

\date{\today}

\begin{abstract}
We consider a noncommutative theory developed in a curved background.
We show that the Moyal product has to be conveniently modified and,
consequently, some of its old properties are lost compared with the
flat case. We also address the question of diffeomorphism symmetry.
\end{abstract}


\maketitle

\section{Introduction}
Usual noncommutative theories \cite{Seiberg} are based on a constant
antisymmetric quantity of rank two. Consequently, they do not
exihibt the Lorentz symmetry for spacetime dimensions higher than two.
In this way, it does not make much sense to look for noncommutative
theories invariant by general coordinate transformations or even
noncommutative quantum fields in curved background.

\medskip
The question of Lorentz invariance in noncommutative theories has been
recently considered \cite{Carlson}, where the antissymetric tensor is
taken as an independent antisymmetric tensor operator. Our purpose in
the present paper is to consider the main aspects of noncommutative
fields in curved space. Our approach is such that the noncommutativity
does not affect the metric but it does on the geommetry. For
simplicity, we deal with the particular case of $D=2$ where there is a
constant flat limit for the noncommutative antisymmetric tensor. The
only difference for dimensions higher than two is that this limit
should be spacetime dependent. We deal with real scalar fields, but
what will be presented here can be naturally extended for any kind of
fields.

\medskip
Our paper is organized as follows. In Sec. II we present the main
aspects of the problem where we see that it is necessity to adopt a
more general definition for the Moyal product. In Sec. III we discuss
in details the property of commutativity for quadractic terms in the
action which is not kept in curved space, and in Sec. IV we consider
the question of diffeomorphism symmetry. We left Sec. V for some
concluding remarks.

\section{Presenting the problem}
\renewcommand{\theequation}{2.\arabic{equation}}
\setcounter{equation}{0}

Let us first consider the usual action for real scalar fields in
curved
spacetime at $D=2$,

\begin{equation}
S=-\frac{1}{2}\int d^2x\,\sqrt{-g}\,g^{\mu\nu}
\partial_\mu\phi\,\partial_\nu\phi\,.
\label{2.1}
\end{equation}

\noindent
Of course, this theory is invariant under general coordinate
transformation. The origin of this symmetry is related fo the
(first-class) constraints \cite{Dirac}

\begin{equation}
\pi^{\mu\nu}=\frac{\partial{\cal L}}{\partial\dot g_{\mu\nu}}=0\,,
\label{2.2}
\end{equation}

\noindent
where ${\cal L}$ is the Lagrangian density corresponding to the action
(\ref{2.1}). The components of $\pi^{\mu\nu}$ given by Eq. (\ref{2.2})
are three (primary) constraints. Constructing the Dirac Hamiltonian
and imposing that these constraints do not evolve in time, one obtain
two secondary constraints. Repeating this procedure, no more
constraints are obtained and all of them are first-class \cite{Dirac}.
According to Castellani \cite{Castellani}, primary first-class
constraints are related to symmetries of the theory. In the present
case, two of them concern to the general coordinate transformation
(embodied in the diffeomorphism algebra generated by the two secondary
constraints) and the last primary constrains is related to the
conformal (Weyl) symmetry.

\medskip
If one direct follows the general rule of transforming usual theories
in noncommutative ones by replacing product of fields by the Moyal
product \cite{Seiberg}, we would have

\begin{equation}
S=-\frac{1}{2}\int d^2x\,\sqrt{-g}\star g^{\mu\nu}
\partial_\mu\phi\star\partial_\nu\phi\,,
\label{2.3}
\end{equation}

\noindent
where $\star$ means the Moyal product, which for any two fields
$\phi_1$ and $\phi_2$, it is defined as

\begin{equation}
\phi_1\star\phi_2=\exp\,\Bigl(\frac{ik}{2}\,
\epsilon^{\mu\nu}\partial_\mu^x\partial_\nu^y\Bigr)\,
\phi_1(x)\phi_2(y)\Big\vert_{x=y}\,,
\label{2.3a}
\end{equation}

\noindent
and $k$ is a constant with mass dimension minus two and
$\epsilon^{\mu\nu}$ is the usual Levi-Civita tensor (in this paper, we
adopt the convention $\epsilon^{01}=1$).

\medskip
The action (\ref{2.3}) does not exhibit the diffeomorphism symmetry
because the canonical momentum conjugate to $g_{\mu\nu}$ is not a
constraint anymore. Further, notice that the Moyal product must also
acts on the elements of $g=\det g_{\mu\nu}$, making the relation
(\ref{2.3}) meaningless for any consistent use.

\medskip
The natural alternative is to consider the derivative operators of the
Moyal product in curved space as covariant ones. In this way, they do
not affect the spacetime geommetry and the inconsistencies mentioned
above disappear. So, for the same two fields  $\phi_1$ and $\phi_2$,
we make

\begin{equation}
\phi_1\diamond\phi_2=\exp\,\Bigl(\frac{ik}{2\sqrt{-g}}\,
\epsilon^{\mu\nu}D_\mu^xD_\nu^y\Bigr)\,
\phi_1(x)\phi_2(y)\Big\vert_{x=y}\,,
\label{2.4}
\end{equation}

\noindent
where the symbol $\diamond$ is to distinguish from the Moyal product
in the flat case. It is important to emphasize that there is no
problem related to the commutativity of the covariant derivatives
because they act in different spacetime coordinates. Notice that in
curved space, even in two dimensions, the characterist antisymmetric
tensor $\theta_{\mu\nu}$ is spacetime dependent, $\theta_{\mu\nu}(x)=
\sqrt{-g}\,\epsilon_{\mu\nu}$ (for the contravariant notation,
$\theta^{\mu\nu}(x)=\epsilon^{\mu\nu}/\sqrt{-g}$).

\medskip
Considering the Moyal product given by (\ref{2.4}), it does not affect
products involving the metric tensor. Consequently, instead of the
action (\ref{2.3}) we simply have

\begin{equation}
S=-\frac{1}{2}\int d^2x\,\sqrt{-g}\,g^{\mu\nu}
\partial_\mu\phi\diamond\partial_\nu\phi\,.
\label{2.5}
\end{equation}

\noindent
However, the price paid is that some usual properties of the Moyal
product are lost. For example, it affects quadratic terms of the
action (we are going to analyse this point with details in the next
section). It is easy to see that the same occurs for cyclic
properties.

\section{A fundamental characteristic of the new Moyal product}
\renewcommand{\theequation}{3.\arabic{equation}}
\setcounter{equation}{0}

The expansion of the exponential operator in (\ref{2.4}) leads to

\begin{eqnarray}
&&\phi_1\diamond\phi_2=\phi_1\phi_2
+\frac{ik}{2\sqrt{-g}}\,
\epsilon^{\mu\nu}\partial_\mu\phi_1\partial_\nu\phi_2
\nonumber\\
&&\phantom{\phi_1\diamond\phi_2=\phi_1\phi_2}
+\frac{1}{2!}\Bigl(\frac{ik}{2\sqrt{-g}}\Bigr)^2
\,\epsilon^{\mu\nu}\epsilon^{\rho\lambda}
D_\rho\partial_\mu\phi_1D_\lambda\partial_\nu\phi_2
\nonumber\\
&&\phantom{\phi_1\star\phi_2=\phi_1\phi_2}
+\dots
\label{3.1}
\end{eqnarray}

\noindent
Under integration over $d^2x \sqrt{-g}$ we directly see that the
second term is zero, but for the third one we have (we drop the
initial factor $(ik/2)^2/2!$)

\begin{eqnarray}
&&\int d^2x\,\sqrt{-g}\,\frac{\epsilon^{\mu\nu}}{\sqrt{-g}}
\frac{\epsilon^{\rho\lambda}}{\sqrt{-g}}\,
D_\rho\partial_\mu\phi_1D_\lambda\partial_\nu\phi_2
\nonumber\\
&&\phantom{\int}
=\int d^2x\sqrt{-g}\bigl(g^{\mu\lambda}g^{\nu\rho}
-g^{\mu\rho}g^{\nu\lambda}\bigr)
D_\rho\partial_\mu\phi_1D_\lambda\partial_\nu\phi_2,
\nonumber\\
&&\phantom{\int}
=-\int d^2x\,\sqrt{-g}\,[D_\mu,D_\nu]\,
\partial^\mu\phi_1\partial^\nu\phi_2\,,
\nonumber\\
&&\phantom{\int}
=-\int d^2x\,\sqrt{-g}\,g^{\mu\rho}
R_{\lambda\rho\mu\nu}\partial^\lambda\phi_1\partial^\nu\phi_2\,,
\nonumber\\
&&\phantom{\int}
=-\int d^2x\,\sqrt{-g}\,R\,g^{\mu\nu}\,
\partial_\mu\phi_1\partial_\nu\phi_2\,,
\label{3.2}
\end{eqnarray}

\noindent
where we have used some identitities listed in the Appendix A. We
observe that by virtue of the curvature this term is not zero. The
same occurs for the next terms of the expansion (\ref{3.1}). Let us
write some of them below (also dropping the factors involving $k$)

\begin{widetext}

\begin{equation}
\int d^2x\,\epsilon^{\mu\nu}
\frac{\epsilon^{\rho\lambda}}{\sqrt{-g}}
\frac{\epsilon^{\eta\xi}}{\sqrt{-g}}
D_\eta D_\rho\partial_\mu\phi_1
D_\xi D_\lambda\partial_\nu\phi_2
=\frac{1}{2}\int d^2x\,\epsilon^{\mu\nu}\Bigl(
\partial_\mu R\,D_\nu\partial_\lambda\phi_1\partial^\lambda\phi_2
+\frac{1}{2}R^2\partial_\mu\phi_1\partial_\nu\phi_2\Bigr)\,,
\label{3.3}
\end{equation}

\begin{eqnarray}
&&\int d^2x\,\epsilon^{\mu\nu}
\frac{\epsilon^{\rho\lambda}}{\sqrt{-g}}
\frac{\epsilon^{\eta\xi}}{\sqrt{-g}}
\frac{\epsilon^{\theta\gamma}}{\sqrt{-g}}
D_\theta D_\eta D_\rho\partial_\mu\phi_1
D_\gamma D_\xi D_\lambda\partial_\nu\phi_2
\nonumber\\
&&\phantom{\int d^2x\,\sqrt{-g}}
=\frac{1}{2}\int d^2x\,\sqrt{-g}\,R\,
\Bigl(3D_\mu\Box\phi_1D^\mu\Box\phi_2
-3D_\alpha D_\mu\partial_\nu\phi_1D^\alpha D^\mu\partial^\nu\phi_2
\nonumber\\
&&\phantom{\int d^2x\,\sqrt{-g}=\int d^2x\sqrt{-g}R}
-RD_\mu\Box\phi_1\partial^\mu\phi_2
-R\partial_\mu\phi_1D^\mu\Box\phi_2
+R^2\partial_\mu\phi_1\partial^\mu\phi_2\Bigr)\,,
\label{3.4}
\end{eqnarray}

\noindent
where $\Box$ is the Laplace-Beltrami operator $\Box=g^{\mu\nu}D_\mu
D_\nu$.

\end{widetext}

\section{Diffeomorphism symmetry}
\renewcommand{\theequation}{4.\arabic{equation}}
\setcounter{equation}{0}

An interesting question concerns to symmetries. It is well-known that
noncommutative theories exhibit the unitarity problem \cite{Gomis}.
This is so due to the presence of higher derivatives. But the problem
here is more subtle because de number of higher derivatives increases
indefinitely. This means that the unitarity problem is accompanied by
a more fundamental question related to the dynamics of these theories.
This is so because dynamics is given by the last order of its time
derivative (in lower orders take place momenta and constraints). If
there is an infinite number of time derivatives, there is an ambiguity
in the limit of these two regions.

\medskip
A way of understanding this problem in a deeper way consists in
analyzing separately the terms of the expansion (isolately, there is
no problem related to the ambiguity of dynamics and constraints)
\cite{Barc1}. Since there is dynamics for the terms of (\ref{3.1}),
that is simpler than the corresponding ones of (\ref{2.5}), let us
consider the second term of the expansion of expression (\ref{3.1}).
Denoting it by $S^{(2)}$, and taking $\phi_1=\phi_2=\phi$, we have

\begin{equation}
S^{(2)}=\frac{1}{2}\int d^2x\,\epsilon^{\mu\nu}
\frac{\epsilon^{\rho\lambda}}{\sqrt{-g}}\,
D_\rho\partial_\mu\phi D_\lambda\partial_\nu\phi\,.
\label{4.1}
\end{equation}

\noindent
Of course, expression (\ref{4.1}) can be written in terms of the
scalar curvature as given in (\ref{3.2}). However, for our purposes
here, it is better to keep it in the way it is. Let us calculate the
momenta. For convenience, we take  them as surface terms lying on a
hypersurface orthogonal to the time direction \cite{Landau}

\begin{equation}
\bar\delta S^{(2)}=\int_{t_0}^tdt\int dx\,\epsilon^{\mu\nu}
\frac{\epsilon^{\rho\lambda}}{\sqrt{-g}}\,
D_\rho\partial_\mu\phi\,
\bar\delta\bigl(D_\lambda\partial_\nu\phi\bigr)\,,
\label{4.2}
\end{equation}

\noindent
where $\bar\delta$ means variation of fields under derivatives
(because just these terms will contribute to the momenta). In this
way, we have for delta term which appears in the equation above

\begin{eqnarray}
\bar\delta\bigl(D_\lambda\partial_\nu\phi\bigr)
&=&D_\lambda\partial_\nu\bar\delta\phi
-\bar\delta\Gamma^\alpha_{\lambda\nu}\partial_\alpha\phi
\nonumber\\
&=&D_\lambda\partial_\nu\delta\phi
-g^{\alpha\beta}\bigl(\partial_\nu\bar\delta g_{\lambda\beta}
-\frac{1}{2}\partial_\beta\bar\delta g_{\lambda\nu}\bigr)
\partial_\alpha\phi,
\nonumber\\
\label{4.3}
\end{eqnarray}

\noindent
where in the last step it was used the symmetry between $\lambda$ and
$\nu$ indices, given in (\ref{4.2}). Developing in a separate way the
terms which will appear after replacing (\ref{4.3}) into (\ref{4.2}),
we have

\begin{eqnarray}
&&\int_{t_0}^tdt\int dx\,\epsilon^{\mu\nu}
\frac{\epsilon^{\rho\lambda}}{\sqrt{-g}}\,
D_\rho\partial_\mu\phi D_\lambda\partial_\nu\bar\delta\phi
\nonumber\\
&&\phantom{\int}
=\int_{t_0}^tdt\int dx\,\Bigl\{
\partial_\lambda\Bigl(\epsilon^{\mu\nu}
\frac{\epsilon^{\rho\lambda}}{\sqrt{-g}}\,
D_\rho\partial_\mu\phi\partial_\nu\bar\delta\phi\Bigr)
\nonumber\\
&&\phantom{\int_{t_0}^tdt\int dx\,}
-\partial_\nu\Bigl(\epsilon^{\mu\nu}
\frac{\epsilon^{\rho\lambda}}{\sqrt{-g}}\,
D_\lambda D_\rho\partial_\mu\phi\bar\delta\phi\Bigr)\Bigr]\,,
\nonumber\\
&&\phantom{\int}
=\int dx\Bigl\{\Bigl[
\partial_1\Bigl(\frac{1}{\sqrt{-g}}D_1\partial_0\phi\Bigr)
+\frac{\epsilon^{\rho\lambda}}{\sqrt{-g}}
D_\lambda D_\rho\partial_1\phi\Bigr]\bar\delta\phi
\nonumber\\
&&\phantom{\int_{t_0}^tdt}
+\frac{1}{\sqrt{-g}}D_1\partial_1\phi\bar\delta\dot\phi\Bigr\}\,.
\label{4.3a}
\end{eqnarray}

\noindent
For the first term involving $\bar\delta g_{\alpha\beta}$, we obtain

\begin{eqnarray}
&&\int_{t_0}^tdt\int dx\,
\frac{\epsilon^{\mu\nu}\epsilon^{\rho\lambda}}{\sqrt{-g}}
D_\rho\partial_\mu\phi\,g^{\alpha\beta}\partial_\nu
\bar\delta g_{\lambda\beta}\partial_\alpha\phi
\nonumber\\
&&\phantom{\int_{t_0}^t}
=\int_{t_0}^tdt\int dx\,\partial_\nu\Bigl(
\frac{\epsilon^{\mu\nu}\epsilon^{\rho\lambda}}{\sqrt{-g}}
D_\rho\partial_\mu\phi\,g^{\alpha\beta}\partial_\alpha\phi
\bar\delta g_{\lambda\beta}\Bigr)\,,
\nonumber\\
&&\phantom{\int_{t_0}^t}
=-\int dx\,\frac{\epsilon^{\rho\mu}}{\sqrt{-g}}D_\rho\partial_1\phi
g^{\alpha\nu}\partial_\alpha\phi\bar\delta g_{\mu\nu}\,.
\label{4.4}
\end{eqnarray}

\noindent
Similarly, for the last term the result is

\begin{eqnarray}
&&\int_{t_0}^tdt\int dx\,
\frac{\epsilon^{\mu\nu}\epsilon^{\rho\lambda}}{\sqrt{-g}}
D_\rho\partial_\mu\phi\,g^{\alpha\beta}\partial_\alpha\phi
\partial_\beta\bar\delta g_{\lambda\nu}
\nonumber\\
&&\phantom{\int_{t_0}^tdt}
=\int dx\,\frac{\epsilon^{\lambda\nu}\epsilon^{\rho\mu}}{\sqrt{-g}}
D_\rho\partial_\lambda\phi
g^{\alpha0}\partial_\alpha\phi\bar\delta g_{\mu\nu}\,.
\label{4.5}
\end{eqnarray}

Replacing (\ref{4.3a})-(\ref{4.5}) into (\ref{4.2}) we identify the
momenta conjugate to $\phi$, $\dot\phi$ and $g_{\mu\nu}$ (which are
the coefficients of the variations of the corresponding fields).
Denoting them respectively by $p$, $p^{(1)}$ and $\pi^{\mu\nu}$, we
have

\begin{eqnarray}
&&p=\partial_1\Bigl(\frac{1}{\sqrt{-g}}D_1\partial_0\phi\Bigr)
-\frac{\epsilon^{\mu\nu}}{\sqrt{-g}}D_\mu D_\nu\partial_1\phi\,,
\label{4.6}\\
&&p^{(1)}=\frac{1}{\sqrt{-g}}D_1\partial_1\phi\,,
\label{4.7}\\
&&\pi^{\mu\nu}=\frac{1}{2\sqrt{-g}}\partial_\alpha\phi
\Bigl(\epsilon^{\rho\mu}g^{\alpha\nu}D_\rho\partial_1\phi
\nonumber\\
&&\phantom{\pi^{\mu\nu}=}
+\epsilon^{\rho\nu}g^{\alpha\mu}D_\rho\partial_1\phi
+\epsilon^{\lambda\nu}\epsilon^{\rho\mu}g^{\alpha0}
D_\rho\partial_\lambda\phi\Bigr)\,.
\label{4.8}
\end{eqnarray}

Notice that $\pi^{\mu\nu}$ is symmetric in $\mu$ and $\nu$ indices, as
it should be. However, the identification of the primary first-class
constraints (responsible for the symmetries of the theory) is not so
apparent as it was in the case of the action (\ref{2.1}), where the
constraints are simple given by (\ref{2.2}). Let us explicitly write
down the components of $\pi^{\mu\nu}$.

\begin{eqnarray}
\pi^{01}&=&-\frac{1}{2\sqrt{-g}}g^{1\alpha}\partial_\alpha\phi\,
D_1\partial_1\phi\,,
\label{4.9}\\
\pi^{00}&=&-\frac{1}{2\sqrt{-g}}g^{0\alpha}\partial_\alpha\phi\,
D_1\partial_1\phi\,,
\label{4.10}\\
\pi^{11}&=&\frac{1}{2\sqrt{-g}}\partial_\alpha\phi
\bigl(2g^{\alpha1}D_0\partial_1\phi
+g^{\alpha0}D_0\partial_0\phi\bigr).
\label{4.11}
\end{eqnarray}

\noindent
If one combine (\ref{4.7}), (\ref{4.9}) and (\ref{4.10}), we observe
that $\pi^{01}$ and $\pi^{00}$ are actually constraints ($\pi^{11}$ is
not). Denoting these constraints by $\psi$ and $\chi$ we have

\begin{eqnarray}
&&\psi=\pi^{01}+\frac{1}{2}g^{1\alpha}\partial_\alpha\phi\,p^{(1)}
\approx0\,,
\label{4.12}\\
&&\chi=\pi^{00}+\frac{1}{2}g^{0\alpha}\partial_\alpha\phi\,p^{(1)}
\approx0\,,
\label{4.13}
\end{eqnarray}

\noindent
where the symbol $\approx$ means weakly zero \cite{Dirac}.
Using the fundamental Poisson brackets

\begin{eqnarray}
&&\{\dot\phi(x),p^{(1)}(y)\}=\delta(x-y)\,,
\label{4.14}\\
&&\{g_{\mu\nu}(x),\pi^{\rho\lambda}\}
=\frac{1}{2}\bigl(\delta_\mu^\rho\delta_\nu^\lambda
+\delta_\mu^\lambda\delta_\nu^\rho\bigr)\delta(x-y)
\label{4.15}
\end{eqnarray}

\noindent
and also

\begin{equation}
\{g^{\mu\nu}(x),\pi^{\rho\lambda}\}
=-\frac{1}{2}\bigl(g^{\mu\rho}g^{\nu\lambda}
+g^{\mu\lambda}g^{\nu\rho}\bigr)\delta(x-y)\,,
\label{4.16}
\end{equation}

\noindent
obtained from (\ref{4.15}), we can actually show that $\psi$ and
$\chi$ are primary first-class constraints, which are related to the
diffeomorphism symmetry. In the case of action (\ref{2.1}), there was
one extra primary first-class constraints responsible for the
conformal symmetry. In the present case, this symmetry is missing.

\medskip
The next step of the present procedure would be the obtainment of
secondary constraints (what has  to be done in a Hamiltonian formalism
keeping velocities \cite{Barc2}) and the construction of the
diffeomorphism algebra. This is just a question of a hard algebraic
work, but we belive that nothing new will appear and the
diffeomorphism algebra should be verified, since the theory is
invariant by general coordinate transformation.

\section{Conclusion}

In this paper we have studied noncommutative fields in curved space.
We consider a formulation where the noncommutativity does not affect
the metric but it does on the geommetry. This procedure
might be taken as an intermediary step to obtain a noncommutative
gravitational theory without inconsistencies on the flat space.

\medskip
We have deal with a spacetime dimensions $D=2$, where the flat limit
of the noncommutative parameter is a constant tensor. Of course, this
could have been presented for higher dimensions. We envisage two
independent formulations for such theories. One of them is to consider
the parameter $\theta_{\mu\nu}$ depending on $x$. In this way, it
would be some antisymmetric tensor field. Another possibility,
followed by Carlson et al. \cite{Carlson}, is to consider
$\theta_{\mu\nu}$ as an independent spacetime coordinate \cite{Barc3}.

\begin{acknowledgments}
This work is supported in part by Conselho Nacional de Desenvolvimento
Cient\'{\i}fico e Tecnol\'ogico - CNPq (Brazilian Research Agency).
One of us, J.B.-N. has also the support of PRONEX 66.2002/1998-9. I am
in debt with Professor R. Amorim for many helpful discussions.
\end{acknowledgments}

\appendix
\section {Useful relations in curved space}
\renewcommand{\theequation}{A.\arabic{equation}}
\setcounter{equation}{0}

For some vector and antisymmetric tensor field ($V^\mu$ and
$F^{\mu\nu}$), we have

\begin{equation}
D_\mu\bigl(\sqrt{-g}\,V^\mu\bigr)
=\partial_\mu\bigl(\sqrt{-g}\,V^\mu\bigr)\,,
\label{A.1}
\end{equation}

\begin{equation}
D_\nu\bigl(\sqrt{-g}\,F^{\mu\nu}\bigr)
=\partial_\nu\bigl(\sqrt{-g}\,F^{\mu\nu}\bigr)\,.
\label{A.2}
\end{equation}

In the particular case of $D=2$, the following relations are true

\begin{equation}
\frac{\epsilon^{\mu\nu}}{\sqrt{-g}}\,
\frac{\epsilon^{\rho\lambda}}{\sqrt{-g}}
=g^{\mu\lambda}g^{\nu\rho}-g^{\mu\rho}g^{\nu\lambda}\,,
\label{A.3}
\end{equation}

\begin{equation}
R_{\mu\nu\rho\lambda}
=\bigl(g_{\mu\rho}g_{\nu\lambda}
-g_{\mu\lambda}g_{\nu\rho}\bigr)\,\frac{R}{2}\,,
\label{A.4}
\end{equation}

\begin{equation}
R_{\mu\nu}=g_{\mu\nu}\,\frac{R}{2}\,,
\label{A.5}
\end{equation}

\end{document}